\documentclass[%
 reprint,
]{revtex4-2}

\usepackage{graphicx}
\usepackage{dcolumn}
\usepackage{bm}
\usepackage{amsfonts}
\usepackage{amsmath}
\usepackage{float}
\usepackage{gensymb}
\usepackage{color}
\newcommand{\ii}{\mathrm{i}}
\newcommand{\ee}{\mathrm{e}}
\usepackage[utf8]{inputenc}
\usepackage{hyperref}
\begin{document}

\title{Bound States to the Continuum: Time-varying Spoof Acoustic Surface Waves}

\author{
E. Paul$^{1}$}
\email{ep564@exeter.ac.uk}
\author{G. J. Chaplain$^{1}$, J. Li$^{1}$, T. A. Starkey$^{1}$,  S. A. R. Horsley$^{1}$
}

\affiliation{$^{1}$Centre for Metamaterial Research and Innovation, Department of Physics and Astronomy, University of Exeter, EX4 4QL, United Kingdom}

\begin{abstract}
\noindent We develop a theoretical framework for time-modulated acoustic metasurfaces comprising a line array of modulated cavities, and show that bound acoustic surface waves can undergo temporal diffraction from bound states localised at an interface into bulk waves. The dispersion relation is derived via an operator formalism that captures the spatio-temporal coupling between Floquet sidebands. We show that under periodic modulation of the cavity length sidebands spaced by the modulation frequency are produced (diffraction in time), enabling the coupling of bound surface acoustic waves with bulk radiation i.e. from a bound state \textit{to} the continuum. We observe the negative-frequency spectra as spatial reflections along the array via time-domain finite element simulations. Spectral $k$-gaps are observed at band crossings, with the width of the gap proportional to the modulation amplitude. The modulation enters solely through a time-dependent reflection phase, such that the framework applies generally to metasurfaces with programmable boundary conditions, beyond purely mechanical modulation.

\end{abstract}
\maketitle

\section{Introduction}

\noindent Shaping wave propagation through time-dependent material parameters has the potential to be a powerful design paradigm across a range of wave systems. Recent focus has been on the periodic modulation of material properties in time -- such as refractive index, impedance, or stiffness -- enabling wave dynamics unattainable in static structures including frequency conversion, directional filtering, and nonreciprocity \cite{shaltout2015time,barati2022optical,fleury2016floquet}. Although theoretical treatments of such media date back several decades \cite{morgenthaler1958velocity}, recent advances in tunable materials and real-time modulation techniques, including magnetic-bias-driven acoustic nonreciprocity \cite{fleury2014sound}, have made such systems practically accessible.

Many of these advances have been realised in electromagnetism, particularly with ultrafast tunable materials such as transparent conducting oxides, e.g. indium tin oxide, which exhibit epsilon-near-zero behaviour, enabling drastic and ultrafast refractive index modulation \cite{kinsey2015epsilon, bohn2021spatiotemporal, shi2024enhanced, alam2016large}. The ability to modulate refractive index with high contrast on femtosecond timescales has enabled a range of phenomena including broadband frequency shifting and angularly tunable temporal refraction \cite{galiffi2024optical}.

In contrast to electromagnetic systems, the implementation of time-varying acoustic properties typically relies on active control, since directly modulating bulk material parameters such as density and bulk modulus is experimentally challenging -- although the theory is well developed \cite{li2019nonreciprocal,zhu2023effective}. 
Recent theoretical work has also explored spatio-temporal modulation of surface admittance for controlling diffusion by a flat surface \cite{kang2025tunable}.
Instead, modulation is most readily introduced at interfaces where the boundary conditions themselves are varied in time \cite{darche2025scattering}, or the effective resonant profiles of individual meta-atoms, often achieved through active feedback \cite{wong2025eigenmode,shen2019nonreciprocal}. Such routes to modulation include sensor–actuator feedback loops, piezoelectric elements, and
digitised speaker-microphone pairs \cite{cho2020digitally,Wang2025}.

Beyond bulk wave manipulation, structured surfaces can support confined surface modes that decay evanescently into the surrounding media. In electromagnetism, surface plasmon polaritons (SPPs) exist, without any surface structuring, as non-radiating charge oscillations at a metal-dielectric interface \cite{barnes2003surface}. At lower frequencies however, where metals behave as perfect electrical conductors, such confined modes are screened. The existence of so-called ``spoof'' SPPs was identified in \cite{pendry2004mimicking} and experimentally developed in \cite{hibbins2005experimental}, using surface corrugations to mimic the field penetration of optical SPPs at GHz frequencies.  Here the dispersion is tailored through geometry \cite{goubau1950surface,gao2018spoof}. This nucleated a breadth of research in metasurfaces for the control of bound surface waves

A key feature of such surface waves is that, at a given frequency, they possess a larger in-plane momentum than bulk (free propagating) modes. As a result, they do not radiate into the surrounding medium and cannot be excited by plane-waves incident from the bulk. At a flat interface, this appears as a mismatch between the in-plane wavevector of the incident bulk wave and that of the surface-bound mode. Introducing periodicity replaces continuous translational symmetry with discrete symmetry, allowing additional in-plane momentum to be supplied in integer multiples of the reciprocal lattice vector. This enables coupling between bulk waves and otherwise inaccessible surface-bound modes, \cite{galiffi2020wood}, a commonly exploited feature of grating couplers \cite{chuang2012physics}. This momentum matching does not by itself render the surface mode radiative; the mode remains non-radiative as long as it lies below the sound line.

The idea of using subwavelength structuring to localise waves at surfaces finds a direct analogue in acoustics, where structured surfaces can support acoustic surface waves (ASWs) \cite{Torrent2012,Jankovic2021a,moore2023acoustic,Starkey2024}. Early experimental work demonstrated ultrasonic surface waves supported by rectangular-groove gratings on rigid substrates \cite{kelders1998ultrasonic}. These resonators (cavities) couple through diffractive near-fields, supporting modes that are confined to the surface and decay evanescently into the surrounding fluid. This same principle of momentum mismatch applies to acoustic surface waves, where periodic structuring not only supports confined modes but also facilitates their excitation from bulk sound waves. Time-varying metasurfaces can facilitate this coupling to and from in an analogous way to spatial diffraction through frequency conversion \cite{galiffi2020wood,chen2021efficient, stefanini2024time}. In this paper we formalise this analytically for an acoustic metasurface with time-varying cavities; their resonant frequencies are modulated through their length.

Traditional tools in the modelling of acoustics, including the renowned transfer matrix method (TMM) \cite{jimenez2021transfer}, have been used recently to model similar spatio-temporally modulated acoustic systems. The expected Doppler-like frequency shifts and modulation-induced modifications to the band structure, including $k$-gaps are found, including associated non-reciprocity in reflection and transmission \cite{cidlinsky2025time}. However, the validity of these models is geometry dependent, as they assume one-dimensional wave propagation and therefore require operation below the first transverse cut-off frequency. Above this cut-off, higher-order modes become propagative and the one-dimensional approximation breaks down As a result, analytically describing metasurfaces that couple surface-bound modes to the surrounding bulk is not possible within this framework.

Instead, we leverage recent advances in an operator-based description of time-varying media \cite{horsley2023eigenpulses}, and derive an effective surface impedance that includes modal-matching at the cavity apertures to the far field. In such operator-based descriptions, the system’s response is represented as a mapping between frequency components of the field, providing a frequency-domain analogue to traditional time-domain approaches. The result is a robust analytical model that reproduces results from time-domain finite element (FE) simulations. We glean physical insight into the conditions under which bound ASW transition to the radiative continuum and predict not only the onset and direction of bulk radiation, but also how it depends on key system parameters such as modulation depth and frequency.

\section{Theory}
\begin{figure}[t]
    \centering
    \includegraphics[width=\linewidth]{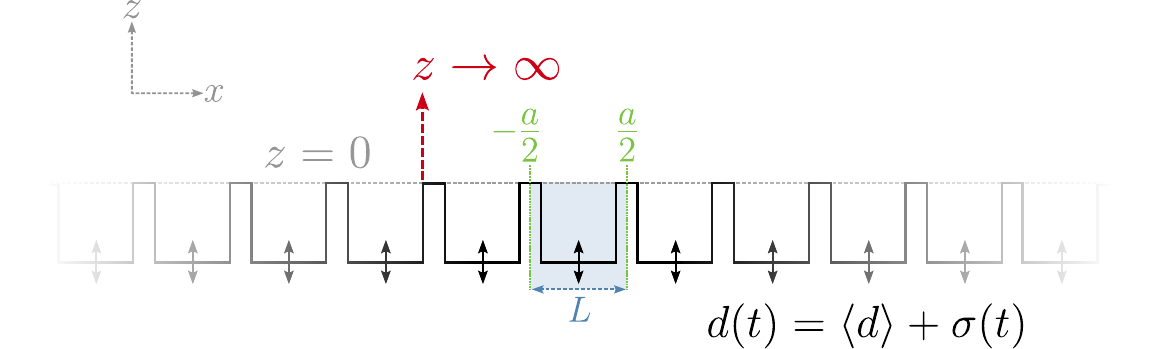}
    \caption{Schematic of the time-modulated acoustic metasurface.
        An infinite one-dimensional array of identical, subwavelength cavities of width $a$ and time-dependent depth $d(t)$ is with cavity openings at $z=0$. The blue highlighted region depicts one unit cell of length $L$. The aperture spans $x\in[-a/2,a/2]$ within each period $L$.
        Arrows denote the modulation of the cavity depth, $d(t) = \langle d\rangle + \sigma(t)$, where $\sigma(t)$ is a small periodic displacement applied in-phase to all cavities.
        Acoustic waves propagate in the half-space $z>0$ above the metasurface.
    }
    \label{fig:geom}
\end{figure}

\noindent We study the propagation of acoustic waves above a periodically modulated metasurface. The surface consists of a one-dimensional array of identical subwavelength cavities, of width $a$ and length $d(t)$ that oscillate sinusoidally in time, and together in-phase; the boundary is therefore periodic in space (through the lattice) and in time (through the modulation). The metasurface lies in the $x$–$z$ plane (invariant in $y$), with grating having period $L$, corresponding to a reciprocal lattice vector of length $g = 2\pi/L$, and periodic in time with period $T$, giving a modulation frequency $\Omega = 2\pi/T$. These two periodicities give rise to Floquet harmonics in both space and time: spatial harmonics and temporal harmonics are indexed by the integers $n,m$ respectively. A schematic is shown in Fig.\ref{fig:geom}.

The field is described by the velocity potential $\varphi(x,z,t)$, which satisfies the linear wave equation
\begin{equation}\label{eq:WE}
\boldsymbol{\nabla}^2\varphi - \frac{1}{c^2}\frac{\partial^2\varphi}{\partial t^2}=0,
\end{equation}
with $\vec v = \boldsymbol{\nabla}\varphi$ and $p=-\rho_0\partial_t\varphi$, prescribing the convention on time-dependence, with $\rho_0$ the fluid density.

We consider wave propagation in the region $z>0$ above the structured surface with the cavities open into the unbounded fluid domain at $z=0$. All boundaries are taken as sound-hard (Neumann), such that the normal velocity in the fluid vanishes.

We now exploit a field expansion of the potential and using the relevant boundary conditions will derive an effective surface impedance operator.

\subsection{Field expansion and boundary conditions}

\noindent The acoustic velocity potential $\varphi(x,z,t)$ satisfies~\eqref{eq:WE} and, owing to the spatial and temporal periodicity of the surface, can be expanded in Floquet harmonics as:
\begin{equation}
    \varphi(x,z,t)=\sum_{n,m} a_{nm}
    \ee^{\ii[(K+ng)x-(\omega+m\Omega)t]}\ee^{-\kappa_{nm}z},
    \label{eq:bloch_expansion}
\end{equation}
where $K$ is the Bloch wavenumber, $\Omega=2\pi/T$ the temporal modulation frequency, and
$\kappa_{nm}^2=(K+ng)^2-(\omega+m\Omega)^2/c^2$ ensures exponential decay for $z>0$. For evanescent surface-bound components, $\kappa_{nm}$ is real and positive, which occurs when $|K+ng| > (\omega+m\Omega)/c$; otherwise $\kappa_{nm}$ becomes imaginary and the corresponding component radiates into the bulk.
At this stage $K$ and $\omega$ are not yet constrained; they simply parameterise the Floquet expansion.  The subsequent boundary conditions will impose a condition that selects the admissible pairs $(K,\omega)$, i.e. the dispersion relation of the surface mode.

To determine the modal amplitudes $a_{nm}$ we apply the boundary conditions at $z=0$.  
Differentiating~\eqref{eq:bloch_expansion} gives the normal velocity
\begin{align}
\begin{split}
    v_z(x,0,t)
    &= \left.\frac{\partial \varphi}{\partial z}\right|_{z=0} \\
    &= -\sum_{n,m} \kappa_{nm} a_{nm} 
      \ee^{\ii[(K+ng)x - (\omega+m\Omega)t]}.
    \label{eq:vz_expansion}
    \end{split}
\end{align}

Along the line $z=0$, at the boundary between the cavities and the homogeneous fluid half-space above, the normal fluid velocity is non-zero only over the apertures and vanishes elsewhere:
\begin{align}
    v_z(x,0,t) = 
    \begin{cases}
    \sum\limits_m v_m \ee^{-\ii(\omega+m\Omega)t}, \quad-a/2\leq x\leq a/2 \\
    0 , \quad \text{otherwise}
    \end{cases}
    \label{eq:v_aperture}
\end{align}
where $v_m$ is the complex amplitude of the $m^{\text{th}}$ temporal Floquet harmonic,
corresponding to the frequency component $\omega + m\Omega$ (hereafter referred to as a
sideband). We denote the sum here by $v_{\text{aperture}}(t)$.

Using standard orthogonality relations over one unit cell, the projection
integrals reduce to the aperture region ($x\in[-a/2,a/2]$), since the
vertical velocity is zero elsewhere. This gives the modal amplitudes
\begin{equation}
    a_{nm}
    = -\frac{v_m}{\kappa_{nm}}\frac{a}{L}
        \mathrm{sinc} \left[\frac{(K+ng)a}{2}\right].
    \label{eq:a_nm}
\end{equation}
Having matched the normal velocity at the aperture openings, we now apply continuity of the velocity potential.

At each aperture, the potential is assumed to be approximately uniform owing to the subwavelength aperture width $a \ll \lambda$. In analogy with the velocity expansion in~\eqref{eq:v_aperture}, we therefore write the
surface potential at $z=0$ as a temporal Fourier series,
\begin{equation}
  \varphi(x,0,t)=\sum_m \varphi_m\ee^{-\ii(\omega+m\Omega)t},
  \label{eq:phi_boundary}
\end{equation}
where $\varphi_m$ is the complex amplitude of the $m^{\text{th}}$ temporal
sideband of the potential evaluated inside the aperture. Within the
aperture $x\in[-a/2,a/2]$, the surface potential must equal this (uniform)
aperture potential. Averaging the boundary value of~$\varphi$ over the
aperture width and substituting the modal amplitudes
from~\eqref{eq:a_nm} gives
\begin{equation}
    -v_m \sum_n 
    \frac{a}{L}
    \frac{\mathrm{sinc}^2 \left[\frac{(K + ng)a}{2}\right]}{\kappa_{nm}}
    = \varphi_m,
    \label{eq:phi_vm_full}
\end{equation}
where we define the geometric overlap factor
\newline $\beta_n=(a/L)\;\mathrm{sinc}^2\big[(K+ng)a/2\big]$, so that
\begin{equation}
  -v_m\sum_n\frac{\beta_n}{\kappa_{nm}}=\varphi_m.
\end{equation}
We also introduce the average inverse decay constant
\begin{equation}
   \langle\kappa^{-1}\rangle_m=\sum_n \frac{\beta_n}{\kappa_{nm}},
   \label{eq:kappa_avg}
\end{equation}
which represents the radiation impedance for sideband $m$. Therefore
\begin{equation}
    -\langle \kappa^{-1} \rangle_m  v_m = \varphi_m.
    \label{eq:final_phi_vm}
\end{equation}
Equation~\eqref{eq:final_phi_vm} links the aperture velocity and potential for each sideband $m$ and forms the basis of the dispersion relation.

In general the velocity potential at the aperture will be linearly related to the velocity by the acoustic impedance of the cavity.  In our case, because time modulation couples different temporal harmonics, the velocity potential at sideband $m$ generally depends on aperture velocities at all sidebands $p$. We capture this with a surface impedance matrix, defined as follows,
\begin{equation}
    \varphi_m = \sum_p Z_{mp} v_p,
    \label{eq:Z_definition}
\end{equation}
where $m$ and $p$ index temporal sidebands.
The diagonal terms $Z_{mm}$ describe the self-response of the $m^{\text{th}}$ sideband, while the off-diagonal terms $Z_{mp}$ ($m \neq p$) capture sideband coupling induced by the modulation. 

We finally substitute \eqref{eq:Z_definition} into \eqref{eq:final_phi_vm}, rewrite both sides as a summation and therefore arrive at

\begin{equation}
\begin{aligned}
    & \sum_p \left[ Z_{mp} + \delta_{mp} \left\langle \kappa^{-1} \right\rangle_m \right] v_p = 0, \\
    &\implies
    \det\left[ Z_{mp} + \delta_{mp} \left\langle \kappa^{-1} \right\rangle_m \right] = 0.
\end{aligned}
\label{eq:dispersion_relation}
\end{equation}
This defines the dispersion relation for surface modes supported by the time-modulated metasurface; the resulting eigenmodes are not confined to a single frequency but span a set of coupled temporal sidebands for a given wavenumber.
In the following section we specify the form of the impedance operator $Z_{mp}$, which encapsulates the local cavity dynamics and the modulation–induced coupling between temporal sidebands.

\subsection{The impedance operator}
\noindent We now derive an expression for the impedance operator $Z_{mp}$ introduced in~\eqref{eq:Z_definition}, which relates the temporal Fourier components of velocity potential and aperture velocity at the surface. This operator characterises the local acoustic response of the individual cavities, including sideband coupling induced by time modulation.

The cavity aperture lies at $z=0$, with the interior extending into $z<0$. A plane wave inside the cavity at frequency $\omega$ has a phase factor of the form $\ee^{-\ii(\omega t \pm k z)} = \ee^{-\ii\omega (t \mp z/c)}$.

As the cavity depth is modulated in time, reflection at the termination couples different frequency components. The field inside the cavity can be written as a superposition of downward-propagating incident sidebands and upward-propagating reflected sidebands. 
We therefore represent the reflection from the cavity termination by a matrix $\boldsymbol{r}$ with elements $r_{mp}$ (whose form depends on the specific modulation), which gives the conversion from an incident sideband $p$ to a reflected sideband $m$. The total cavity potential is then
\begin{equation}
\begin{aligned}
\varphi_{\mathrm{cav}}(z,t)
&= \sum_{p} A_p \ee^{-\ii(\omega+p\Omega)\left(t-\frac{z}{c}\right)} \\
&\quad+ \sum_{m,p} r_{mp} A_p \ee^{-\ii(\omega+m\Omega)\left(t+\frac{z}{c}\right)}.
\end{aligned}
\label{eq:cavity_field}
\end{equation}
where $A_p$ are the incident downward amplitudes at each sideband.
As expected, in the static limit the matrix is diagonal, $r_{mp}\to r_{m}\delta_{mp}$, whereas with temporal modulation, the off-diagonal entries are populated. For example, \cite{hooper2025harnessing} describes the experimental measurement of the reflection matrix in an electromagnetic time-varying medium at GHz frequencies.

Differentiating \eqref{eq:cavity_field} gives the corresponding vertical velocity:
\begin{equation}
\begin{aligned}
v_z(z,t)
    &= \ii \sum_p \frac{\omega + p\Omega}{c} A_p
        \ee^{-\ii(\omega + p\Omega) \left(t - \frac{z}{c}\right)} \\
    &\quad - \ii \sum_{m,p} \frac{\omega + m\Omega}{c} r_{mp} A_p
        \ee^{-\ii(\omega + m\Omega) \left(t + \frac{z}{c}\right)}.
\end{aligned}
\label{eq:vz_full}
\end{equation}

Evaluating at the cavity opening ($z=0$) and once again using orthogonality relations gives, 
\begin{equation}
    v_m = \ii\frac{\omega + m\Omega}{c}
          \left[-A_m + \sum_p r_{mp} A_p \right].
    \label{eq:v_m_full}
\end{equation}
The first term here represents the directly incident contribution at sideband $m$, and the second accounts for coupling from all other sidebands through the reflection matrix.

Similarly, $\varphi_{\text{cav}}$ becomes
\begin{equation}
    \varphi_m = A_m + \sum_p r_{mp} A_p,
    \label{eq:phi_m_full}
\end{equation}
which defines the complex velocity potential at each temporal sideband.

Because these relations couple all sidebands simultaneously, it is
convenient to rewrite them in vectorised form, where the algebra leading
to the impedance operator becomes more compact. We therefore collect the sideband amplitudes into vectors.
Let $\mathbf{A} = (\ldots, A_{-1}, A_0, A_1, \ldots)^\top$ collect the
incident amplitudes and let $\mathbf{r}$ denote the reflection matrix with
elements $r_{mp}$. Strictly speaking, $\mathbf{r}$ is an infinite-dimensional
operator, but in practice we truncate the system to a finite number of
sidebands (e.g.\ $m,p=-M,\ldots,M$), which is sufficient provided the
modulation is weak enough that higher-order coupling becomes negligible.
From \eqref{eq:phi_m_full} and \eqref{eq:v_m_full} we obtain
\begin{equation}
    \boldsymbol{\varphi} = (\mathbf{1} + \mathbf{r})\cdot\mathbf{A}, 
    \qquad
    \mathbf{v} = -\ii\boldsymbol{\Lambda}\cdot(\mathbf{1} - \mathbf{r})\cdot\mathbf{A},
\end{equation}
where $\boldsymbol{\Lambda}={\rm diag}[(\omega+m\Omega)/c]$, i.e.
$\Lambda_{mp} = \delta_{mp}(\omega+m\Omega)/c$.
Eliminating the vector $\mathbf{A}$ gives a direct relation between the vector of frequency components of the velocity potential, $\boldsymbol{\varphi}$, and the vector for the frequency components of the velocity, $\mathbf{v}$,
\begin{equation}
    \boldsymbol{\varphi}
    = \ii(\mathbf{1}+\mathbf{r})\cdot(\mathbf{1}-\mathbf{r})^{-1}\cdot
      \boldsymbol{\Lambda}^{-1}\cdot\mathbf{v},
      \end{equation}
from which we identify the impedance operator
\begin{equation}
    \mathbf{Z}
      = \ii(\mathbf{1}+\mathbf{r})\cdot(\mathbf{1}-\mathbf{r})^{-1}\cdot
        \boldsymbol{\Lambda}^{-1},
    \label{eq:Z_operator}
\end{equation}
whose elements $Z_{mp}$ are those introduced in~\eqref{eq:Z_definition}. To solve the full dispersion relation, all that is now required is to evaluate the form of the reflection operator, which we do here for a simple sinusoidal modulation.

\subsection{The reflection operator}
We model the time-dependent cavity depth as
\begin{equation}
    d(t) = \langle d\rangle + \sigma(t), \qquad |\sigma(t)| \ll \langle d\rangle,
    \label{eq:depth_def}
\end{equation}
where $\sigma(t)$ denotes a small amplitude modulation about the mean depth $\langle d\rangle$.  
The coordinate $z$ is measured upward from the aperture ($z=0$), so the rigid cavity wall lies at
    $z = -d(t) = -\langle d\rangle - \sigma(t).$

The normal fluid velocity at the wall must equal the wall velocity. Taking $+z$ upward, the boundary condition is therefore $v_z = -\dot{\sigma}(t)$.

We express the velocity field within the cavity 
in terms of the transformed times
$\tau_\mp = t \mp z/c$ \cite{cidlinsky2025time}, such that the moving wall
\begin{equation}
    \tau_-^\star(t) = t + \frac{\langle d\rangle + \sigma(t)}{c},
    \qquad
    \tau_+^\star(t) = t - \frac{\langle d\rangle + \sigma(t)}{c}.
    \label{eq:tau_at_wall}
\end{equation}

\noindent We note that the wall motion must remain subsonic, i.e.\ $|\dot{\sigma}(t)| < c$. 
This ensures that the transformed time arguments $\tau_\pm^\star(t)$ in
\eqref{eq:tau_at_wall} remain monotonically increasing functions of $t$, so that the
mapping between $t$ and $\tau_\pm^\star$ is one–to–one.  
This monotonicity is required in order to invert the transformation in
\eqref{eq:tau_at_wall} and apply the orthogonality of exponentials in the projected
boundary condition.

Incorporating the position of the moving–wall and applying the sound-hard boundary condition at it
gives
\begin{equation}
\begin{aligned}
&\ii\sum_{p}\frac{\omega+p\Omega}{c}  A_p  
    \ee^{-\ii(\omega+p\Omega)\tau_-^\star(t)}  \\
&\quad
-  \ii\sum_{m,p}\frac{\omega+m\Omega}{c}  r_{mp}A_p  
    \ee^{-\ii(\omega+m\Omega)\tau_+^\star(t)}
= -\dot\sigma(t).
\end{aligned}
\label{eq:wall_bc_expanded}
\end{equation}

For subsonic, small wall displacements the term $-\dot\sigma(t)$ represents weak direct radiation from the wall (that is essentially a moving piston), that is neglected at leading order. This contribution has no fixed phase relationship with the incident field, simply acting as an additional independent acoustic source that does not participate in the coupling between temporal sidebands induced by the time-varying cavity depth. In practice such piston–type radiation could be isolated or subtracted experimentally, so we omit it here and retain only the modulation-induced coupling.

We once again use the orthogonality of exponentials (with the transformed time) to retrieve the set of coupled relations between incident and reflected
sidebands,
\begin{equation}
\frac{\omega+p\Omega}{c}  A_p
= \sum_{m,p'} \frac{\omega+m\Omega}{c}  r_{mp'}A_{p'}  I_{mp},
\label{eq:proj_balance_general}
\end{equation}
where we define
\begin{equation}
I_{mp}=\frac{1}{T}\int_0^T
\ee^{-\ii[(\omega+m\Omega)\tau_+^\star(t)-(\omega+p\Omega)\tau_-^\star(t)]}dt.
\label{eq:I_def}
\end{equation}
This is a time-averaged overlap integral accounting for phase modulation at the moving wall.

For excitation containing a single incident sideband $p$
(i.e. all other $A_{p'}=0$ for $p'\neq p$),
the coupled relation~\eqref{eq:proj_balance_general} simplifies to
\begin{equation}
\sum_{m} (\omega+m\Omega)r_{mp}I_{mp}
= (\omega+p\Omega),
\label{eq:rmp_relation}
\end{equation}
from which each coupling coefficient can be identified.

Substituting the explicit forms of $\tau_\pm^\star(t)$ into~\eqref{eq:I_def}  gives
\begin{equation}
I_{mp} = \ee^{\ii\chi_{mp}\langle d\rangle} 
\frac{1}{T}\int_0^T \ee^{\ii(p-m)\Omega t}\ee^{\ii\chi_{mp}\sigma(t)}dt,
\end{equation}
where
\begin{equation}
\chi_{mp} = \frac{2\omega+(m+p)\Omega}{c}.
\label{eq:chi_def}
\end{equation}
The associated reflection coefficients are then
\begin{equation}
r_{mp}
= \left(\frac{\omega+p\Omega}{\omega+m\Omega}\right)
\frac{\ee^{\ii\chi_{mp}\langle d\rangle}}{T}\int_0^T
\ee^{\ii(p-m)\Omega t}  \ee^{\ii\chi_{mp}\sigma(t)}  dt.
\label{eq:rmp_final}
\end{equation}
Equation \eqref{eq:rmp_final} provides a general expression for the frequency-conversion coefficients induced by a time-dependent cavity boundary.
Importantly, the modulation enters solely through the phase factor $\ee^{\ii\chi_{mp}\sigma(t)}$, demonstrating that sideband coupling arises from a time-dependent reflection phase rather than from an explicit source term. As a result, the theory is not restricted to mechanical modulation but applies equally to any implementation that realises an effective time-dependent phase upon reflection.

We now explicitly assume sinusoidal wall motion, $\sigma(t) = \alpha \sin(\Omega t)$ with $\alpha \ll d$,

and define the dimensionless wall-speed parameter
$e = \alpha \Omega / c$, so that
$1 + \dot{\sigma}(t)/c = 1 + e\cos(\Omega t)$.

The mapping between the time $t$ and the phase variable
$\tau_-^\star(t)=t+\frac{\langle d\rangle+\sigma(t)}{c}$
for a sinusoidal modulation $\sigma(t)=\alpha\sin(\Omega t)$
takes the form
\[
\Omega\tau_-^\star = \Omega t + e\sin(\Omega t),
\]
which is mathematically identical to Kepler’s equation
$M = E - e\sin E$ for an orbit of eccentricity $e$.
The non-uniform relation between $\tau$ and $t$ therefore mirrors
the non-uniform angular motion in an eccentric orbit:
regions of faster and slower apparent phase advance correspond
to the wall’s forward and backward motion, manifesting as eccentricity in time. The inversion of this is well studied, although to leading order the transformation is trivial \cite{cidlinsky2025time}.

The reflection coefficients can therefore be written as
\begin{equation}
\begin{aligned}
r_{mp}
= &\left(\frac{\omega + p\Omega}{\omega + m\Omega}\right)
\ee^{\ii\chi_{mp}\langle d\rangle}
\frac{1}{T}\int_0^T [1+e\cos(\Omega t)] \\[0.4em]
&\times \ee^{\ii(p-m)\Omega t}  \ee^{\ii q_{mp}\sin(\Omega t)}  dt,
\label{eq:rmp_mod}
\end{aligned}
\end{equation}
where the modulation index is defined as
\begin{equation*}
q_{mp} = \chi_{mp}\alpha
= \frac{2\omega + (m+p)\Omega}{c}\alpha
= 2\frac{\alpha\omega}{c} + e(m+p).
\label{eq:qmp_def}
\end{equation*}

Using the Jacobi–Anger expansion and averaging over one period, to leading order in $\alpha$,
only three terms in the harmonic sum contribute:
\[
s = m-p, \qquad s = m-p-1, \qquad s = m-p+1.
\]
Substituting this result into~\eqref{eq:rmp_mod} gives
\begin{equation}
\begin{aligned}
r_{mp}
= &\left( \frac{\omega + p\Omega}{\omega + m\Omega} \right)
\ee^{\frac{2\ii}{c}\left(\omega + \tfrac{1}{2}(m+p)\Omega\right)\langle d\rangle} \\
&\times \Bigl[
J_{m-p}(q_{mp})
+\tfrac{e}{2}J_{m-p-1}(q_{mp})
+\tfrac{e}{2}J_{m-p+1}(q_{mp})
\Bigr],
\label{eq:rmp_final_mod}
\end{aligned}
\end{equation}
which gives the complete frequency–coupling relation for a sinusoidally modulated cavity depth.

Using the reflection matrix $r_{mp}$ from~\eqref{eq:rmp_final_mod} in the impedance-operator definition~\eqref{eq:Z_operator} yields the impedance matrix $\mathbf Z$.  
Once the modulation is defined -- here through its periodic form $\sigma(t)$ -- the corresponding coefficients $r_{mp}$ fully determine the impedance operator and hence the metasurface dispersion.  
The supported modes are obtained from the non-trivial solutions of \eqref{eq:dispersion_relation}, which we evaluate numerically in the following section by truncating to a finite set of sidebands $m\in\{-M,\dots,M\}$.

\section{Results and discussion}

\noindent We evaluate the eigenvalue condition of Eq.~\eqref{eq:dispersion_relation} by computing the smallest eigenvalue of the matrix 
  $[Z_{mp} + \delta_{mp}\langle\kappa^{-1}\rangle_m]  $.  
When this eigenvalue approaches zero, the determinant of the matrix vanishes and a surface mode is supported.

\begin{figure*}[t]
  \centering
  \includegraphics[width=\linewidth]{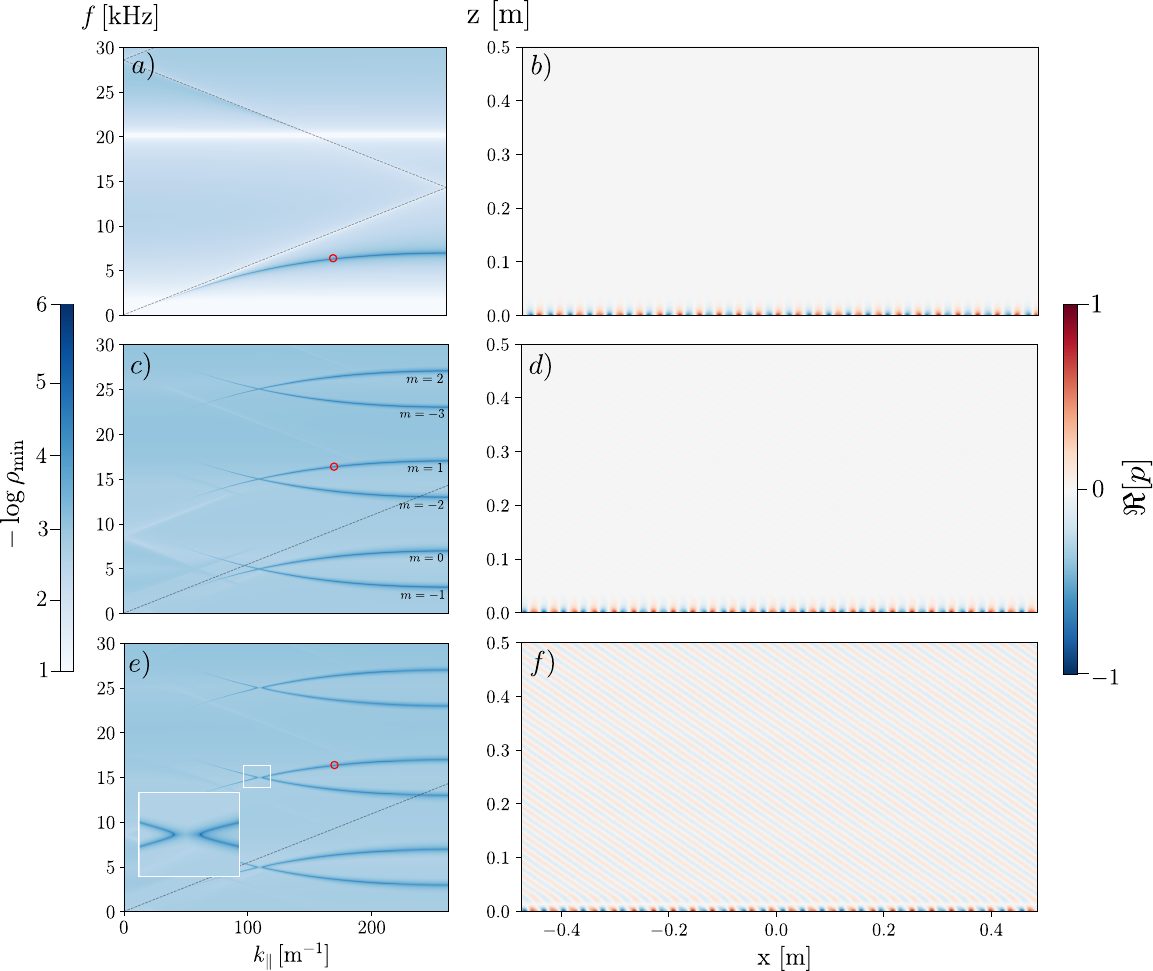}
  \caption{Numerical solution to analytical dispersion relation and reconstructed real pressure fields: (a,b) Unmodulated metasurface showing the acoustic surface wave (ASW) branch below the sound line (black dotted line) and its exponentially decaying pressure field, evaluated at $k_\parallel = 170~\mathrm{m^{-1}}$ and $f = 6.37$~kHz.
  (c,d) Weak modulation ($\alpha = 0.01d$, $f_\text{mod}=10$~kHz) generates Floquet sidebands spaced by the modulation frequency. 
  Negative-index branches ($m<0$) are folded upward into the positive spectrum; the red circle marks the representative point ($k_\parallel = 170~\mathrm{m^{-1}}$, $f = 16.37$~kHz) used for the corresponding field plot. Sideband indices are labelled in (c).
  The pressure field shows a superposition of the evanescent ASW and a very faint radiative component.
  (e,f) Strong modulation ($\alpha = 0.1d$) enhances sideband coupling, producing hybridisation between reflected branches ($m'=-m-1$) and opening modulation-induced band gaps in $k_\parallel$ (inset), the size of which scales with the modulation amplitude. The red circle in (e) marks the same representative point ($k_\parallel = 170~\mathrm{m^{-1}}$, $f = 16.37$~kHz) used in (c), at which the corresponding field in (f) is reconstructed.
  The pressure field in (f) exhibits pronounced radiation into the bulk, the emission angle of which depends on $k_\parallel$, while the surface-bound envelope remains visible near the interface.
  }
\label{fig:big}
\end{figure*}

Figure~\ref{fig:big} summarises the analytical results for three representative cases:  
(a,b) the unmodulated metasurface, serving as a reference i.e. $\alpha = 0$;  
(c,d) the weakly modulated case with $\alpha = 0.01d$;  
and (e,f) a strongly modulated case with $\alpha = 0.1d$.  
Here $\alpha$ denotes the amplitude of the cavity–depth modulation
$\sigma(t)=\alpha\sin(\Omega t)$ introduced in Eq.~\eqref{eq:depth_def}, so that
the instantaneous cavity depth is  
$d(t)=\langle d\rangle + \sigma(t)$.
The left column shows the dispersion maps, plotting the logarithm of the smallest eigenvalue ($-\log\rho_{\min}$). Supported modes appear as dark bands in the spectrum, while features such as the bulk sound line or cavity resonances produce only shallow minima and remain bright; divergence of the impedance results in half-wavelength resonances, that physically do not satisfy the Neumann boundary conditions; zeros in $\kappa$ correspond to radiative modes. The real pressure fields $\Re(p)$ evaluated at the points indicated by the red circles are shown in the respective rows of the right column.

\paragraph*{Static surface - Fig.~\ref{fig:big}(a,b)}
The metasurface consists of an infinite array of cavities of depth $d = 8.5$ mm, aperture width $a = 8$ mm, and unit-cell width $L = 12$ mm. The corresponding resonance frequency is $f_0 = 7$ kHz. 
The dashed line in Fig.~\ref{fig:big}(a) represents the sound-line $f = c k_0 / 2\pi$ i.e. the acoustic dispersion of the bulk fluid above the metasurface. The component of the wavevector parallel to the surface, $k_{\parallel}$, is related to the free-space wavenumber $k_0$ by $k_\parallel=k_0\sin\theta$ for incidence angle $\theta$; the sound line corresponds to grazing incidence.

The numerical results demonstrate the expected zone-folding behaviour, with the sound line `folding back' (diffracting) at the Brillouin-zone edge $k_\parallel=\pi/L$. 
The horizontal band around $f \simeq 20$ kHz corresponds to a standing-wave resonance of the cavities. 
By contrast, the guided solutions of the metasurface correspond to values of $(k_\parallel,\omega)$ for which the determinant approaches zero, giving rise to the dark band below the sound line that identifies the acoustic surface wave (ASW) branch.
In this regime the vertical decay constant $\kappa_{nm}$ is real and positive, so the field decays exponentially into the fluid half-space and remains bound to the surface. This condition is met when $k_\parallel>k_0$: for a given frequency $\omega$, a bulk sound wave in free space satisfies the classical dispersion relation $k_\parallel^2+k_\perp^2 = (\omega/c)^2$, with $k_\perp$ being the component of the wavevector perpendicular to the surface. The sound line therefore marks the boundary between radiating and non-radiating solutions: modes lying below the line require an in-plane wavenumber $k_\parallel$ larger than that of a free plane wave at the same frequency. This excess momentum forces the out-of-plane component $k_\perp$ to become imaginary, corresponding to exponential decay away from the surface. Such modes are confined to the interface and appear in the map as the ASW band. 


To visualise the localisation of the surface wave to the metasurface, we evaluate the field at the point marked by the red circle in Fig.~\ref{fig:big}(a), corresponding to
$k_\parallel = 170~\mathrm{m^{-1}}$ and $f = 6.37$ kHz on the ASW branch. The acoustic pressure is obtained from the velocity potential $\varphi(x,z,t)$ from the standard definition, using the derived form of $\varphi$. 
The resulting real part of the normalised pressure amplitude, shown in Fig.~\ref{fig:big}(b), confirms the expected localisation: the field is confined to the surface and decays exponentially into the bulk. We plot the field over 80 unit cells, matching the numerical domain used in Finite Element (FE) analysis later.

\paragraph*{Weak modulation - Fig.~\ref{fig:big}(c,d)}

Introducing temporal modulation alters this picture. A sinusoidal modulation of the cavity depth at frequency $f_\text{mod}$ generates Floquet sidebands at $\omega+m\Omega$, where $m \in \mathbb{Z}$.
In the following we set $f_\text{mod}=10$~kHz. 

For the case shown in Fig.~\ref{fig:big}(c,d), the modulation follows $\sigma(t)=\alpha\sin(\Omega t)$ with displacement amplitude $\alpha=8.5\times10^{-5}$~m, corresponding to 1\% of the cavity depth ($d=8.5$~mm). The associated dimensionless wall-speed parameter is $e=\alpha\Omega/c\approx0.02$, which measures the ratio of the wall’s maximum velocity to the sound speed. This confirms that the modulation remains subsonic, consistent with the assumptions of the theory. In the dispersion plots that follow, the $m=0$ branch represents the original ASW band, while sidebands with $m\neq 0$ appear as shifted replicas with $10$kHz spacing, including the $m=-1, -2, -3$ branches that `fold up' from negative frequencies.

In the dispersion map of Fig.~\ref{fig:big}(c), the $m=0$ branch corresponds to the original ASW band. A sequence of sidebands is generated, each offset vertically by the modulation frequency $f_\text{mod}=10$~kHz. Negative-index branches ($m<0$) are reflected upward into the positive-frequency spectrum, so
that in this particular frequency window the $m=-1$, $m=-2$, and $m=-3$ sidebands are visible. These sideband indices are labelled in Fig.~\ref{fig:big}(c). The analytics formally involve an infinite set of
sidebands, but the contribution of higher orders decays rapidly with $|m|$. This follows from the Bessel-function dependence of $r_{mp}$ in Eq.~\eqref{eq:rmp_final_mod}.
Consequently, we focus only on the first two
sidebands carrying significant amplitude, consistent with the finite set resolved in the FE simulations (shown below). We therefore truncate the expansion to
$m\in[-M,M]$; here $M=3$ captures all features.

The red circle marks the same in-plane wavenumber ($k_\parallel = 170~\mathrm{m^{-1}}  $) used in the static case, allowing direct comparison with the pressure field shown in Fig.~\ref{fig:big}(d).  
Although the envelope remains surface-bound, temporal sidebands at the same in-plane wavenumber $k_\parallel$ lie above the sound line; these components therefore possess a real out-of-plane wavevector and give rise to radiative leakage into the bulk, which is faintly visible in the field.

\paragraph*{Strong modulation - Fig.~\ref{fig:big}(e,f)}
Increasing the modulation amplitude by an order of magnitude to $\alpha = 8.5 \times 10^{-4}$~m (10\% of the cavity depth) amplifies sideband coupling and enables stronger hybridisation between Floquet branches.  
At this modulation amplitude, crossings between reflected pairs $m' = -m - 1$ (e.g. $m = 0$ with $m' = -1$, $m = 1$ with $m' = -2$) hybridise where they meet, opening modulation-induced band gaps in $k_\parallel$ resulting from the interference between positive and negative frequencies \cite{hendry2025effects}. 

Since this modulation is temporal rather than spatial, the system mixes frequency components and does not conserve energy, whereas the in-plane momentum $k_\parallel$ remains conserved; the hybridisation therefore lifts degeneracies along $k_\parallel$, leading to the observed momentum-space band-gaps.
These appear clearly in Fig.~\ref{fig:big}(e), with an inset highlighting one such gap.  
When two sidebands of frequencies $\omega + m\Omega$ and $\omega + m'\Omega  $ intersect in $\omega$-$k_\parallel$ space, the time-periodic modulation provides an additional coupling term oscillating at $\Omega$, lifting the degeneracy and splitting the dispersion locally into two branches separated by a gap whose width scales with $\alpha$. The widening of such gaps appears important for applications, and can be achieved with similar modulation strengths in photonic systems (with time-varying resonances) \cite{wang2025expanding}, or by increasing the overlap through dispersion engineering (i.e. metamaterials raison d'être). 

The corresponding pressure field, shown in Fig.~\ref{fig:big}(f), exhibits a pronounced radiation into the bulk, with angle predictable from the dispersion relation, whilst maintaining a visible surface-bound component near the interface.
\begin{figure*}
  \centering
  \includegraphics[width=\linewidth]{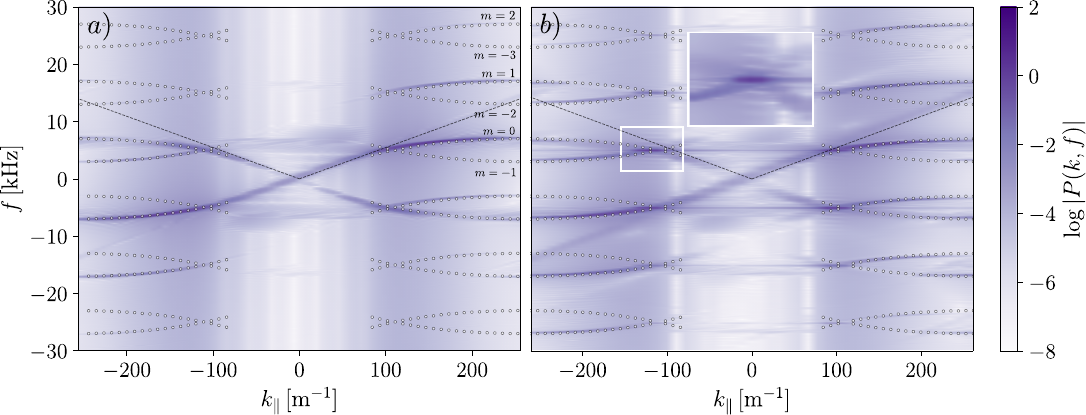}
  \caption{
  Finite-element dispersion maps of the time–modulated cavity metasurface.
  Each panel shows the spatio-temporal Fourier transform of the pressure field
  recorded $10$~mm above a finite array of $80$ cavities, excited by a point
  source at one end of the array. Increasing the modulation amplitude from 1\% (a) to 10\% (b) shows strengthened higher-order sidebands. The same branches as in Fig.~\ref{fig:big}(c) are labelled for reference. Overlaid points indicate the corresponding analytical sideband positions.
  Regions that appear as avoided crossings (momentum-space gaps) in the
  analytical dispersion manifest here with increased Fourier amplitude at the temporal band-edge (inset),
  consistent with parametric amplification in the driven, time-modulated
  system \cite{hooper2025quasi}. The dashed line marks the sound line.}  
  \label{fig:fem}
\end{figure*}

To verify the analytical model, we performed finite-element method (FEM) simulations of a modulated cavity array with the same geometry and modulation parameters.
A finite-duration point source was placed in the leftmost cavity to excite the surface, and the pressure field was recorded along a line 10 mm above the surface.
A two-dimensional Fourier transform of this pressure data with respect to time and in-plane position yields the numerical dispersion spectrum shown in Fig.~\ref{fig:fem}. Table~\ref{tab:fem_params} (Appendix) shows parameters used in the FE model.

Unlike the analytical model, which identifies all supported modes of the infinite system, the FEM spectrum reveals only those modes that are physically excited by the source and efficiently coupled to the far field.
Because the excitation is applied only at one end of the array, the system is not symmetric under $k_\parallel \rightarrow -k_\parallel$.

The point source preferentially launches surface waves travelling in the positive-$k_\parallel$ direction; these waves then reflect from the far end of the finite array, producing a weaker backward-propagating component.
Time modulation mixes this reflected field with the carrier, generating frequency-shifted sidebands that therefore appear more pronounced at negative $k_\parallel$.
Despite this asymmetry, the overlaid analytical sideband positions (dots) align closely with the dominant FEM features, confirming agreement between theory and simulation.

In the analytical dispersion, where hybridisation between Floquet branches appears as avoided crossing, the driven FEM simulations probe the response of an active, time-modulated system. At the sideband intersections, the temporal modulation couples counter-propagating and frequency-shifted components. In the analytical eigenvalue problem, this coupling lifts the degeneracy and opens a momentum-space gap; this formulation identifies the existence and structure of the eigenmodes but does not encode how strongly they are excited or whether they exchange energy with the modulation.

By contrast, the FEM simulations probe a driven response: at the sideband intersections, the temporal modulation coherently couples counter-propagating and frequency-shifted components of the surface wave.
In these regions the coupled components are phase-matched, allowing the time-periodic boundary to perform net work on the field and transfer energy into the acoustic surface mode. The same coupling that opens a gap in the analytical dispersion therefore appears in the FEM as parametric amplification \cite{hooper2025quasi}, leading to an enhanced pressure fields and thus Fourier coefficients in the numerical dispersion maps at the temporal band edge. This effect is visible in the strong-modulation case ($\alpha = 0.1d$) in Fig.~\ref{fig:fem}(b) (highlighted in inset).
The FEM results are subject to finite-size effects arising from both the limited cavity array length (80 unit cells) and the finite simulation time, which constrain the achievable resolution in $k_\parallel$ and frequency. 

\section{Conclusion}

\noindent We have developed a theoretical framework for an acoustic metasurface with time-modulated cavity boundaries, deriving a full dispersion relation that captures spatio-temporal coupling between Floquet sidebands through the impedance and reflection operators. Because the modulation enters only through a time-dependent reflection phase, the formulation applies to general periodic modulation profiles and to systems supporting any number of coupled sidebands.

The model predicts the generation of sidebands spaced by the modulation frequency, with negative-frequency branches reflected upward into the positive spectrum.  
These sidebands couple to bulk radiation modes when they cross above the sound line, producing radiation into the far field at an angle dependent on the modulation frequency.  
At higher modulation amplitudes, interaction between reflected branches opens modulation-induced band gaps in the in-plane wavevector, the width of which scales with the modulation strength.  

Finite-element simulations of a driven, finite cavity array confirm the predicted sideband structure and radiative behaviour. While the analytical model identifies the supported eigenmodes of an infinite system, the numerical simulations probe the response of an actively driven, time-modulated structure. In this driven setting, regions that appear as momentum-space gaps in the analytical dispersion manifest (due to resolution) as amplification of the Fourier coefficients at the temporal band-edge, consistent with parametric amplification enabled by temporal modulation. The close alignment between analytical sideband positions and numerical spectra demonstrates strong agreement between the two approaches.

The analysis done in this work also highlights practical considerations for possible experimental implementation. Appreciable sideband generation and radiative coupling require modulation depths at the percent level, with hybridisation effects emerging for modulation amplitudes approaching $10\%$ of the cavity depth. Percent-level mechanical modulation has been demonstrated experimentally in coupled acoustic resonators, albeit at low modulation frequencies \cite{li2019nonreciprocal}. 

Reaching the stronger modulation regime explored in this analysis remains challenging at kilohertz frequencies, and finite array length and measurement time impose limits on wavevector and frequency resolution in numerical and experimental implementations. Nevertheless, the formulation developed here provides quantitative guidance for possible future experimental design and realisation.

An additional practical advantage of the analytical framework developed here is
its computational efficiency.  
Because the impedance and reflection operators reduce to a truncated matrix
problem involving only a small number of Floquet sidebands -- typically
$m\in{-2,-1,0,1,2}$, which is sufficient to capture all physically relevant
coupling -- the full dispersion relation can be evaluated far more efficiently than
finite-element simulations. This efficiency makes the analytical theory particularly valuable as a design tool, enabling quick exploration of parameter space before committing to large-scale numerical simulations or experimental design.

Finally, while the present work has focused on a mechanically modulated cavity depth, the theoretical framework also applies more generally. The modulation enters through the phase accumulated upon reflection, and the same spatio-temporal coupling can therefore be realised using alternative implementations that impose an effective time-dependent phase response (\eqref{eq:rmp_final}). In practice, this opens routes to non-mechanical realisations, for example via digitally programmed boundary conditions \cite{cho2020digitally} or active acoustic elements that synthesise the required modulation depths without physical motion. Such approaches may offer greater experimental flexibility, particularly at higher frequencies where mechanical modulation becomes challenging.

\acknowledgements

\noindent The authors acknowledge the financial support by the EPSRC (grant no EP/Y015673/1) and would also like to thank Dr I.~R.~ Hooper for helpful conversations. ‘For the purpose of open access, the author has applied a ‘Creative Commons Attribution (CC BY) licence to any Author Accepted Manuscript version arising from this submission’.

\bibliography{bib}

\section*{Appendix: Finite-Element Modelling Details}
 \noindent Here we summarise the finite-element (FEM) procedure used to validate the analytical dispersion model. All simulations were performed in COMSOL Multiphysics 6.2 using a 2D pressure-acoustics formulation, with coupling to a deformed geometry interface to realise time-varying cavity lengths. The model is two-dimensional and consists of a finite array of $80$ identical subwavelength cavities of cavity depth $d$ and aperture width $a$, arranged with lattice period $L$. The cavities radiate into a large half-space above the surface with perfectly matched layers included to simulate the unbounded domain. A summary of all key model parameters is provided in Table~\ref{tab:fem_params}.

The structure is driven by a localised monopole point source placed at the entrance of the leftmost cavity. The source emits a 5-cycle pulse at the carrier frequency $f_0$, ensuring broadband excitation of the surface-wave.
Time modulation is implemented by prescribing a sinusoidal vertical displacement at the cavity bases corresponding to the analytical model. All cavities are modulated in phase with modulation frequency $\Omega$ and amplitude corresponding to either $1\%$ or $10\%$ of the unmodulated cavity depth. The pressure field is recorded along a horizontal line $10$ mm above the apertures and processed via a two-dimensional Fourier transform in space and time to obtain the numerical dispersion maps.

\begin{table}[H]
  \centering
  \caption{Key parameters used in the finite-element simulations.}
  \begin{tabular}{ll}
    \hline
    Parameter & Value \\
    \hline
    Cavity depth & $d = 8.5~\mathrm{mm}$ \\
    Aperture width & $a = 8~\mathrm{mm}$ \\
    Lattice period & $L = 12~\mathrm{mm}$ \\
    Number of unit cells & $N_\text{cells} = 80$ \\
    Carrier frequency & $f_0 = 7~\mathrm{kHz}$ \\
    Modulation frequency & $f_\text{mod} = 10~\mathrm{kHz}$ \\
    Modulation depth (weak) & $\alpha = 0.01 d$ \\
    Modulation depth (strong) & $\alpha = 0.1 d$ \\
    Timestep (1\% modulation) & $\Delta t = \text{1.6e-5}~\mathrm{s}$ \\
    Timestep (10\% modulation) & $\Delta t = \text{8e-6}~\mathrm{s}$ \\
    Total simulation time & $t_\text{tot} = 0.01$ s\\
    \hline
  \end{tabular}
  \label{tab:fem_params}
\end{table}

\end{document}